\documentclass[prl,twocolumn,preprintnumbers,superscriptaddress]{revtex4}
\usepackage{amsmath,amssymb,mathrsfs}
\usepackage{graphicx}
\usepackage{hyperref}
\usepackage{paralist}

\usepackage{subfigure}
\usepackage{dcolumn}
\usepackage{color}
\usepackage{bbold}

\def\be{\begin{equation}}
\def\ee{\end{equation}}
\def\nn{\nonumber}

\begin{document}
\title{Spirals and skyrmions in two dimensional oxide heterostructures} 
\author{Xiaopeng Li}
\affiliation{Department of Physics and Astronomy, University of Pittsburgh, Pittsburgh, Pennsylvania 15260, USA}
\affiliation{Kavli Institute for Theoretical Physics, University of California, Santa Barbara, CA 93106, USA}

\author{W. Vincent Liu}
\affiliation{Department of Physics and Astronomy, University of Pittsburgh,
  Pittsburgh, Pennsylvania 15260, USA}
\affiliation{Center for Cold Atom Physics, Chinese Academy of Sciences, Wuhan 430071, China}

\author{Leon Balents} 
\affiliation{Kavli Institute for Theoretical Physics, University of California, Santa Barbara, CA 93106, USA}

\begin{abstract}

We construct the general free energy governing long-wavelength
magnetism in two-dimensional oxide heterostructures, which applies
irrespective of the microscopic mechanism for magnetism.   This leads,
in the relevant regime of
weak but non-negligible spin-orbit coupling, to a rich phase
diagram containing in-plane ferromagnetic, 
spiral, cone, and skyrmion lattice phases, as well as a nematic state
stabilized by thermal fluctuations.  
\end{abstract}

\maketitle  

\noindent{\bf Introduction.} 
Metallic interfaces between insulating oxides, such as
SrTiO$_3$ (STO)-LaAlO$_3$ (LAO) or STO-GdTiO$_3$ (GTO), provide a
versatile platform to study two dimensional electron liquids.
Numerous experiments have observed
magnetism in such structures~\cite{2007_Brinkman_LAOSTO_NatMat,
  2011_Li_LAOSTO_NPHYS,2011_Dikin_Coexist_SCFM_PRL,2011_Ariando_LAOSTO_Ncomm,
  2011_Bert_LAOSTO_FMSC_NPHYS,2012_Kalisky_LAOSTOFM_ncomms,
2012_Moetakef_STOFM_PRX,2013_Joshua_LAOSTO_PNAS,2013_Bi_FM_arXiv},
which has
generated tremendous interest in emergent many-body phenomena of the
interfacial electrons.  The mechanism behind the magnetism is
presently controversial.  Theory has not yet even reached a consensus
on whether the magnetic moments arise from localized or extended
electrons.  Possible explanations include a charge ordered state of
interfacial
electrons\cite{pentcheva2006charge,banerjee2013ferromagnetic},
ferromagnetism of a single TiO$_2$ layer mediated by RKKY coupling
from extended subbands\cite{2012_PatrickLee_LAOSTO_PRL}, fully
itinerant magnetism\cite{PhysRevLett.110.206401}, local moment
formation assisted by disorder\cite{PhysRevLett.110.206401}, and
oxygen defect states\cite{pavlenko2012oxygen}.  One may even
contemplate extrinsic explanations~\cite{2005_Abraham_NonMag_APL}.

Despite these uncertainties, the macroscopic properties of the magnetization and its
consequences for transport are interesting and require theoretical
understanding.  In LAO-STO, torque magnetometry measurements indicate
in-plane moments with unusual field
dependence~\cite{2011_Li_LAOSTO_NPHYS}, and significant spin-orbit
coupling (SOC) effects are observed~\cite{2010_Caviglia_SO_LAOSTO_PRL,
  2010_Shalom_LAOSTO_SOC_PRL,2012_Fete_SO_LAOSTO_PRB}.  Experiments
have demonstrated remarkable tunability of electronic properties and
phase transitions at the interface~\cite{2008_Caviglia_LAOSTO_nature,
  2011_Joshua-IIani_CriticalDensity,2013_Joshua_LAOSTO_PNAS,2008_Cen_NanoWire_NatMat,2013_Bi_FM_arXiv},
making this a promising platform to study the interplay of
ferromagnetism and SOC.  The obvious fact that the
presence of an interface breaks inversion symmetry suggests the
possibility of analogies to novel helical and skyrmion states studied
intensely recently in 
{non-centrosymmetric materials~\cite{2006_Rosseler_Skyrmion_Nature} such as
MnSi\cite{muhlbauer2009skyrmion}, 
Fe$_{0.5}$Co$_{0.5}$Si\cite{yu2010real} 
and magnetic thin films}~\cite{2001_Bogdanov_MagThinFilms_PRL,2007_Bode_ChiralSurface_Nature,2011_Heinze_ChiralSurface_NPHYS}. 

\begin{figure}[htp]
\includegraphics[angle=0,width=.7\linewidth]{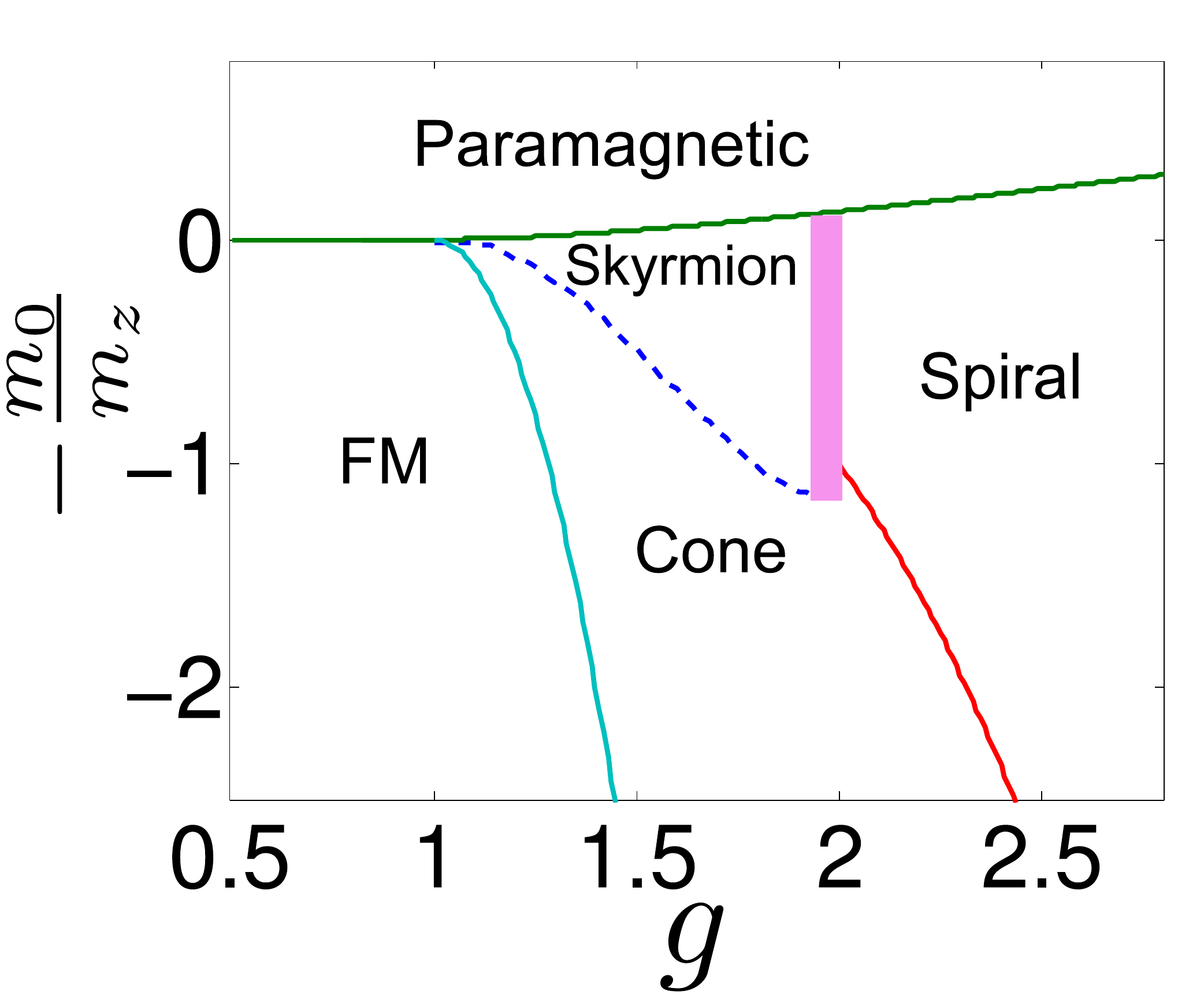}
\caption{Mean field phase diagram parametrized by the coupling $g$ and
  $-\frac{m_0}{m_z}$, which may be taken as a measure of reduced
  temperature  (see text).   On lowering the temperature, the
  paramagnetic phase gives way to in-plane ferromagnetic (FM), spiral,
  cone, and skyrmion lattice phases, as shown.   Solid and dashed
  lines denote continuous and first order transitions, respectively,
  as determined by an analysis of critical phenomena.  The shaded
  region marks a narrow sequence of two transitions, as discussed in
  the text.   } 
\label{fig:mfphasediag}
\end{figure}

In this article, we take a phenomenological approach based on
symmetry, which is valid irrespective of the microscopics.  We assume
only the SOC is weak (in a precise sense formulated below) and
describe the consequences for magnetism.  The generic form of the free
energy is derived, and includes both anisotropy terms and a linear
derivative coupling~\cite{1964_Dzyaloshinskii_SPJETP,1989_Bogdanov_JETP,1994_Bogdanov_JMMMat} 
that plays a central role in driving spin
modulation instabilities.  The general approach is indeed quite
analogous to the theory of the skyrmion lattice states just mentioned.
Minimizing this free energy in the weak SOC regime, we find a rich
phase diagram (Fig.~\ref{fig:mfphasediag}) including in-plane
ferromagnetic (FM), spiral, cone, and skyrmion lattice states.  The
skyrmion lattice state is similar to those discussed above, but has a
staggered arrangement of topological skyrmion charge.

To complete our study of the phase diagram, we discuss fluctuation
effects based upon renormalization group analysis and general
arguments.  We determine the universality classes of the various
transitions, and more interestingly argue that fluctuations generate a
nematic phase between the spiral and paramagnetic ones.  


\noindent{\bf Symmetry analysis and free energy. } 
Without SOC, the phenomenological free energy { (with Landau coefficients $m_0, K$
and $d_0$ in standard notation ~\cite{1980_Lifshitz_StatPhys}) } describing spin fluctuations 
near a ferromagnetic transition in two dimensions is
\begin{eqnarray}
\textstyle 
F_0 = \int d^2 \vec{r}  \left\{ -m_0 | \vec{S}|^2  + 
K\sum_{j=x/y}  |\partial_j \vec{S}|^2  
+ d_0 |\vec{S}|^4 \right\} , 
\label{eq:F0}
\end{eqnarray}
where $\vec{S}(\vec{r})$ is a three component vector field describing
spin density, and we use $j$ to index the two spatial dimensions, with
sums on repeated indices implied in the following.  Here we have
neglected the temporal fluctuations, which is valid at non-zero
temperature away from quantum critical regime.  $F_0$ has SO(3) spin
rotation symmetry. For a square lattice with SOC,
{\it either of atomic or Rashba type,} the symmetry is lower. There is
a space-spin combined lattice rotation symmetry $C_4$, defined as 
\begin{eqnarray}
C_4: \left[\begin{array}{c}
    S_x (\vec{r}) \\
    S_y(\vec{r}) \\
    S_z (\vec{r})
             \end{array}
    \right] 
\to
{\cal R} (\alpha)
   \left[ 	\begin{array}{c} 
                     S_x (R^{-1} (\alpha) \vec{r}) \\ 
		     S_y (R^{-1} (\alpha) \vec{r}) \\
		     S_z (R^{-1} (\alpha) \vec{r}) 
                \end{array}
    \right],  
\end{eqnarray}
with 
$
{\cal R} = \left[ \begin{array}{cc}
            R(\alpha) &0 \\
	    0	&1
          \end{array}
    \right]
$
and 
$R(\alpha) =
\left[
\begin{array}{cc}
 \cos (\alpha) &-\sin(\alpha) \\
  \sin(\alpha) & \cos(\alpha) 
\end{array}
\right], 
$
where $\alpha$ takes the values of $0$, $\pi/2$, $\pi$ or $3 \pi/2$. 
And there is a reflection symmetry $\mathscr{R}$, 
where $\vec{S}$ transforms as a pseudovector 
\begin{eqnarray}
\mathscr{ R} :\left[ \begin{array}{c} 
	S_x (x,y)  \\
	S_y (x,y)  \\ 
	S_z (x,y)  
        \end{array}
	  \right] 
      \to
      \left[ \begin{array}{c} 
	 S_x (-x, y) \\
	 -S_y (-x, y) \\ 
	 -S_z(-x,y) 
        \end{array}
	  \right]  
\end{eqnarray} 
Also we have time-reversal symmetry ${\cal T}: \vec{S} \to -\vec{S}$.

The general form of non-SO(3)-invariant terms in the free energy
consistent with these symmetries is
$
F = F_0 + \Delta F
$ 
with 
\begin{eqnarray}
&&\Delta F = \int d^2 \vec{r} 
    \left\{ m_z S_z ^2 + \lambda S_z \partial_j S_j \right. \nn\\
    &&+ b_1 \left[ (\partial_x S_y) ^2 + (\partial_y S_x) ^2 \right] 
    + b_2 \partial_x S_x \partial_y S_y 
    +b_3 \partial_j S_z \partial_j S_z \nn \\ 
    &&+ \left. d_1 S_z ^ 4 + d_2 S_x ^2 S_y ^2 + d_3 S_z^2 S_j S_j 
      +{\cal O} (\partial^2 S^4, S^6)
  \right\} .
\label{eq:freeenergy},  
\end{eqnarray}
{where $m_z$, $\lambda$, $b_{1,2,3}$ and $d_{1,2,3}$ are 
phenomenological coupling constants~\cite{1980_Lifshitz_StatPhys}. } 
These terms arise from SOC, and we anticipate $\lambda
\propto {\cal O} (\lambda_\mathrm{SO} ) $, $m_z \propto {\cal O}
(\lambda_\mathrm{SO} ^2)$, and $b_1, b_2, b_3 \propto {\cal O}
(\lambda _\mathrm{SO} ^2$), where $\lambda_\mathrm{SO}$ is the
microscopic SOC strength. We verify this explicitly for a         
microscopic model {to be discussed elsewhere}~\cite{Li_Liu_Balents_unpub}.  However, we note that the
theory does not require this scaling, but only that $\lambda, m_z,
b_{1,2,3}$ are small for weak SOC.   Note that $\lambda$ is linear in
derivatives, which suggests it is possible to lower the energy by
forming a state with non-zero wavevector.   This is borne out by more
detailed analysis, as we will see.

\noindent{\bf Mean field phase diagram.} 
Near the spin ordering transition temperature $T_c$ the spin configuration 
is controlled by the quadratic part of the free energy, 
which reads
\begin{eqnarray} 
F^{(2)} = \int \frac{d^2 \vec{q}} {(2\pi)^2} T_{\alpha \beta} (\vec{q}) 
    \tilde{S}_\alpha (\vec{q}) \tilde{S}_\beta (-\vec{q}), 
\label{eq:F2}
\end{eqnarray} 
after a Fourier transformation 
$
S_\alpha (\vec{r}) = \int \frac{d^2 \vec{q}}{(2\pi)^2}
      \tilde{S}_\alpha (\vec{q}) e^{i \vec{q} \cdot \vec{r}}. 
$
The matrix $T$  in Eq.~\eqref{eq:F2} is given by 
\begin{eqnarray}
T (\vec{q}) &=& (-m_0+K \vec{q}^2) \mathbb{1}_{3\times 3} \nn \\
    &+& \left[ 
      \begin{array}{ccc} 
       b_1 q_y ^2 		& \frac{b_2}{2} q_x q_y		& \frac{i\lambda}{2} q_x \\ 
       \frac{b_2}{2} q_x q_y 	& b_1 q_x ^2 			& \frac{i\lambda}{2} q_y \\
      -\frac{i\lambda}{2} q_x   & -\frac{i\lambda}{2} q_y	& m_z + b_3 \vec{q}^2
      \end{array}
\right]. 
\end{eqnarray} 
We assume $m_z>0$, which favors in-plane moments, as typical for 2d
systems due to dipolar effects, and in accordance with experiments on
LAO-STO~\cite{2011_Li_LAOSTO_NPHYS}.  Neglecting for the moment the
$b_{1,2,3}$ terms, we obtain the lowest eigenvalue of the $T$ matrix
as:
\begin{eqnarray} 
\textstyle \epsilon (\vec{q}) = -m_0 + K q^2 + 
\tfrac{1}{2} \left[ m_z- \sqrt{m_z^2 + \lambda^2 q^2 }
\right] + {\cal O} (b q^2).
\label{eq:evalsmallq}  
\end{eqnarray} 
{Introducing the dimensionless ratio }
$$
\textstyle g \equiv \frac{\lambda^2}{4 K   m_z},
$$
 one observes that when $g>1$, $\epsilon(q)$ is indeed minimized by a non-zero wavevector $q=Q$, where
$$Q  = \frac{\lambda}{4K} \frac{\sqrt{g^2-1}}{g}.$$ For $g<1$ the
out-of-plane component of the spiral is too costly and the minimum
remains at $q=0$.  The corresponding eigenvalue is
\begin{equation}
  \label{eq:1}
  \epsilon(Q) = - m_0 - \frac{m_z (g-1)^2}{4g} \Theta(g-1),
\end{equation}
where $\Theta(x)$ is the Heavyside step function.

The couplings $b_{1,2,3}$ can be treated perturbatively, and produce
anisotropy in $\vec{q}$ space.  The leading order correction to the eigenvalue is 
$$\Delta\epsilon(\vec{q}) = \kappa_- b_3 + \kappa_+ (2b_1 + b_2) \frac{q_x^2 q_y^2}{q^4},$$
where $\kappa_\pm = \frac{q^2}{2}(1 \pm m_z/\sqrt{m_z^2 + q^2 \lambda^2}) >0$.  
This favors {\sl axial} spirals with wavevectors $\vec{Q}_1 = (Q,0)$ and $\vec{Q}_2 = (0, Q)$ if $b_1> -\frac{b_2}{2}$, 
and {\sl diagonal} spirals with $\vec{Q}_1 = \frac{1}{\sqrt{2}} ( Q, Q)$ and $\vec{Q}_2 = \frac{1}{\sqrt{2}} ( -Q, Q)$, otherwise.

When $\epsilon(\vec{Q})<0$, the system orders, and we introduce two complex
order parameters $\phi_1, \phi_2$ by writing
\begin{eqnarray} 
\vec{S} (\vec{r})=\sum_{\nu =1,2} \left[ \phi_\nu (\vec{r}) e^{i\vec{Q}_\nu \cdot \vec{r}} \vec{e}_\nu + c.c. \right],  
\label{eq:spiralorder}
\end{eqnarray} 
where $\vec{e}_\nu$ is the eigenvector of $T$ matrix at momentum $\vec{Q}_\nu$. 
For the axial spiral phases, these are 
$\vec{e}_1 ^\mathrm{ \,axial} \approx -i \cos (\varphi/2) \hat{x} + \sin (\varphi/2) \hat{z}$,  
and $\vec{e}_2 ^\mathrm{ \,axial}  \approx -i \cos(\varphi/2) \hat{y} + \sin (\varphi/2) \hat{z}$, 
where $\tan \varphi = \sqrt{g^2-1}$. The eigenvectors for diagonal spiral phases are given by a $\pi/4$ rotation as 
$ 
\vec{e}_\nu ^\mathrm{\, diag}   = {\cal R} (-\frac{\pi}{4}) \vec{e}_\nu ^\mathrm{\, axial}.
$

\begin{figure}[htp]
\includegraphics[bb = 7 7 593 428,angle=0,width=.73\linewidth]{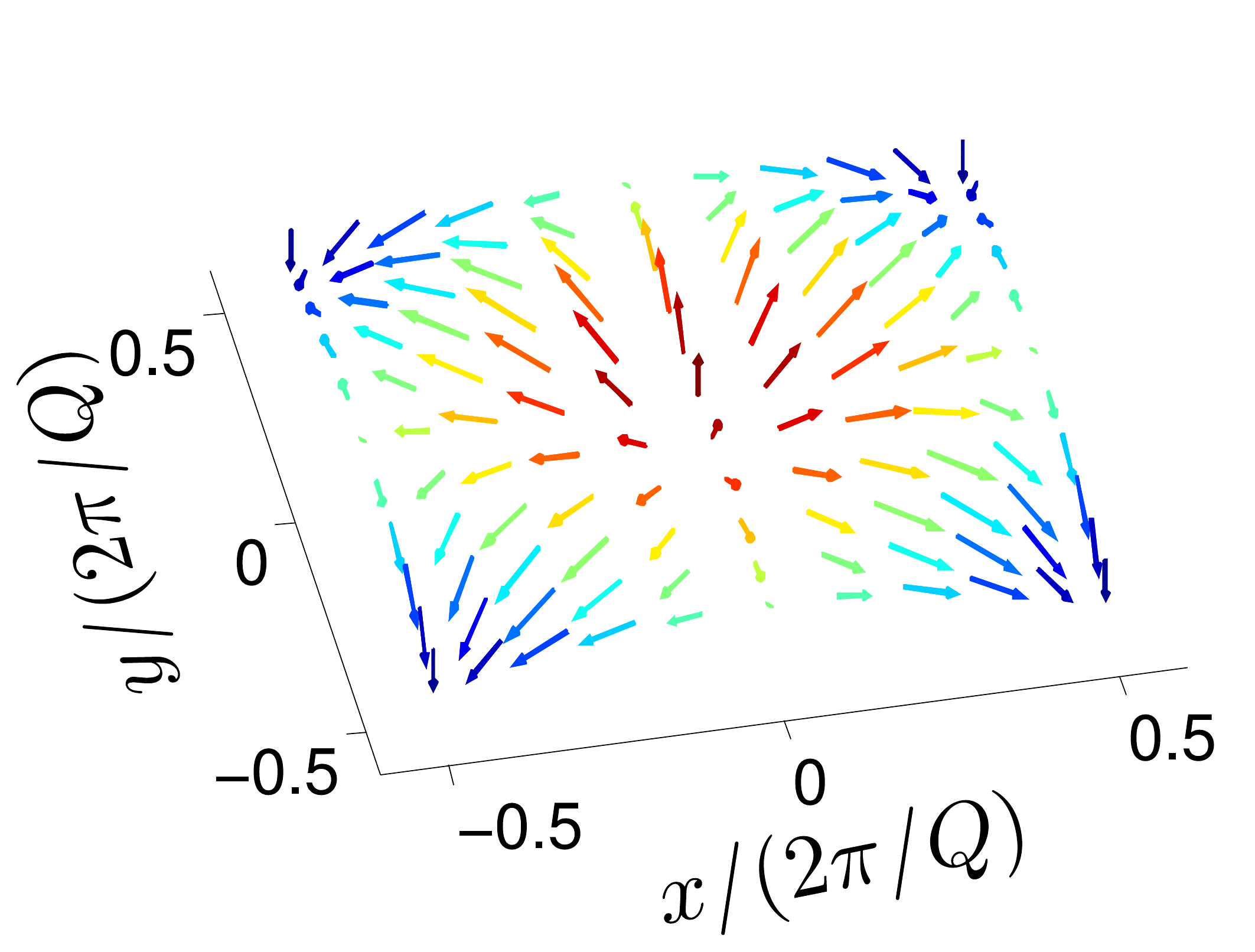}
\caption{
Spin configuration of an axial skyrmion lattice state. 
One unit cell is plotted here. Nodal points where the spin moment vanishes, i.e., $S_x = S_y = S_z =0$, 
are stable in presence of the space-spin combined $C_4$ and time-reversal symmetries.}
\label{fig:spirallattice}
\end{figure}

The above analysis determines the modes involved in the ordering just
below the transition from the paramagnetic state, $\epsilon (Q) =0^-$
(Eq.~\eqref{eq:1}), when the magnitude
of $\vec{S}$ is infinitesimal.   For $g<1$,  this implies in-plane
ferromagnetism, while for $g>1$, there is a degeneracy of states with
different choices of $\phi_\nu$.  This is split by quartic terms in
the free energy, which can be written as
\begin{eqnarray}
\label{eq:spiralenergy}
 f_\mathrm{spiral} \approx &&4 \epsilon (Q) \left(\rho_1  + \rho_2 \right)  
   + 4 d_0 \left\{ \left( 4 + 2 \cos^2 (\varphi) \right) \left(\rho_1 + \rho_2 \right) ^2  \right. \nn \\
	&&        + 4\left(1-2\cos(\varphi) \right) \rho_1 \rho_2 
	\big\} ,
\end{eqnarray}
with $\rho_\nu = \frac{1}{2} |\phi_\nu|^2 $.  Other quartic terms
$d_1$, $d_2$ and $d_3$ are not considered for the reason that they
only provide subleading corrections here.  
From Eq.~\eqref{eq:spiralenergy}, 
spin modulations with two components $|\phi_1| = |\phi_2|$ are favorable if
$\cos (\varphi) >\frac{1}{2}$; otherwise a spiral phase with
($\phi_1=0$, $\phi_2 \neq 0$) or ($\phi_1 \neq 0$, $\phi_2 =0$) is 
favorable. The boundary between these two is at $ g = 2$.  The spin
configuration in { the two-component} $(\phi_1,\phi_2)$ phase is plotted in Fig.~\ref{fig:spirallattice}, 
showing that it can be regarded as a {\em skyrmion lattice}, 
{
which is a crystalline state of spirals with one spin texture per unit cell. 
Different from previous skyrmions discussed for other chiral 
magnets~\cite{2006_Rosseler_Skyrmion_Nature,2009_Yi_Nagaosa_PRB,2010_Han_Nagaosa_PRB}, 
the present skyrmion lattice state has nodal points (Fig.~\ref{fig:spirallattice}), which are 
protected by the space-spin combined $C_4$ and time-reversal symmetries. 
The nodal points can be removed by breaking time-reversal symmetry, 
for example, with an external magnetic field.  }

On lowering temperature further, the competition amongst phases
shifts, with quartic terms playing a more important role.  By a
combination of stability analysis of the above ordered states and
numerical/analytical minimization of the free energy, we establish the
phase diagram in Fig.~\ref{fig:mfphasediag}.   The minimal free energy
states can be approximately expressed in the form
$$
\vec{S}(\vec{r}) = \vec{S}_0 + \sum_{\nu =1,2} \left[ \phi_\nu (\vec{r}) e^{i\vec{Q}_\nu \cdot \vec{r}} \vec{e}_\nu + c.c. \right],
$$ 
where $\vec{S}_0, \phi_\nu, \vec{e}_\nu$ are parameters that vary in
the different phases.  Between FM and spiral phases, we obtain a
coexistence state, with both $\vec{S}_0$ and one $\phi_\nu$ non-zero
(and $\vec{S}_0\cdot \vec{e}_\nu=0$), known as a ``cone'' state. 
{ 
This state arises due to the competition of linear derivative and quartic terms
in the free energy.  }   
The boundaries of the cone state with the FM and spiral phases are given respectively
by $g=1+\sqrt{\frac{2 d_2 m_0}{d_0 m_z}}$ (to leading order in $m_0$)
and $-\frac{m_0}{m_z} = -\frac{1}{2} g(g-1)^2$, and
represent continuous transitions. 
{
The cone state can also be reached from a skyrmion lattice state by 
lowering the temperature.}  The phase transition from
the skyrmion lattice to the cone state is found to be first order.  
An intermediate phase, an anisotropic  two-component spiral state, occurs in a narrow
transition region between the spiral and skyrmion lattice states~(Fig.~\ref{fig:mfphasediag}), 
via a pair of second order transitions.

\noindent{\bf Fluctuation effects.  }
The above mean-field analysis neglects fluctuation effects, which can
be significant in two dimensional systems with continuous broken
symmetry.  Here we discuss them based upon renormalization group and
other arguments.

The in-plane ferromagnetic state,  as well as the uniform component of the
polarization in the coexistence/cone state, is pinned by anisotropy
arising from SOC to lie along one of the easy axes.  Thus it behaves
as a discrete Ising order parameter and long-range order is thereby
stabilized by SOC.  The situation is more subtle for the spiral
states, which have a continuous degeneracy associated with the
phase(s) of $\phi_\nu$.    Note that as a result of spatial
translation symmetry, the problem has an emergent $U(1)\times U(1)$ symmetry: $\phi_\nu \to
\phi_\nu e^{i\theta_\nu} $.  This is true for both axial and diagonal spiral
phases.  

Deep inside the spiral phases, the effect of fluctuations is
understood simply by rewriting $\phi_\nu =
|\phi_\nu|e^{i\vartheta_\nu}$, and considering quadratic fluctuations
of $\vartheta_\nu$.  These are, as usual, logarithmic, and thereby
generate power-law correlations of the non-zero $\phi_\nu$ fields.
Thus the spiral order becomes {\em quasi-}long-range rather than truly
static, though the physics is otherwise largely unchanged.

The transitions between the spiral and other phases are more
interesting.  They can be understood, following the pioneering work of
Kosterlitz and Thouless (KT), from the point of view of unbinding of
topological defects.  This gives rather different results for the two
spiral phases.

Consider first the spiral phase, where only a single spiral order
parameter is non-zero.  The order parameter manifold consists of two
disconnected parts, associated to the two possible wavevectors.  Here
there are two types of topological defects.  The first is a vortex in
the phase of the spiral, which as usual is a point defect with a
logarithmic energy cost 
({see Fig.~S1 in Supplementary Materials}).  The second is a domain wall separating
regions of the two possible spiral wavevectors $\vec{ Q}_1$ and 
$\vec{ Q}_2$.  This is a defect with a non-zero large tension in two
dimensions.  The relatively small energy cost of the vortices is
readily overcome by entropy, and we expect a conventional KT
transition with increasing temperature. 
{ For example, the anomalous dimension at this transition 
is expected to be $1/4$.}

Since the domain walls are not involved in this transition, their line
tension remains positive just above the KT transition, and the system
remains in a single ``wavevector'' domain.  Formally, this means the
order parameter $\langle \rho_1-\rho_2 \rangle \neq 0$, so the $C_4$
rotation symmetry is spontaneously broken, though time-reversal and
translation symmetry is restored.  Fluctuations therefore induce an
intermediate {\em nematic} phase between the spiral and paramagnetic 
phases.  The transition from the nematic to paramagnetic state occurs at a
higher temperature, and is expected to be of Ising type.

Applying the same reasoning, the transition from the cone to FM
phase is also expected to be KT-like.  This is consistent because the
$C_4$ symmetry is broken in the FM by the uniform in-plane moment.

Now consider the skyrmion lattice state. Here the order parameter
manifold is continuous and fully connected.  It is described just by
the angular variables $\vartheta_1$ and $\vartheta_2$, which translate
the lattice in either the x or y direction.  Consequently, the only
defects are vortices in the two phase fields, or equivalently
dislocations of the lattice 
({see Fig.~S2 in Supplementary Materials}).  Therefore only a single transition is
expected.  This may also be guessed from the fact that the skyrmion
lattice preserves $C_4$ symmetry.  The transition should be described
as the melting of this skyrmion lattice, which appears to be in the
same universality class as the melting of a square lattice on a
tetragonal but incommensurate substrate.  Such 2d melting transitions
were analyzed by Halperin and Nelson, and share similar
characteristics with KT
transitions~\cite{1978_Nelson_twodmelting_PRB,1973_KT_JPC}.

{
The above conclusions describe likely critical properties based on
universality from the general theory of 2d critical phenomena, but for
example first order transitions could also occur for non-universal
reasons.  Hence we vetted them against a non-perturbative but
approximate renormalization group (RG)~\cite{2002_Berges_ERG_PHYSReport} 
calculation for spiral states, 
where it is useful to rewrite the free energy (Eqs.~\eqref{eq:F0} and~\eqref{eq:freeenergy}) 
in terms of slowly-varying order parameters $\phi_\nu$ as 
\begin{eqnarray} 
&& F = \textstyle \int d  \vec{r} 
\left\{ 
	\frac{Z}{2}\left[  \left(|\partial_x \phi_1|^2 + |\partial_y \phi_2|^2 \right) \right. \right. \nn \\
 &&+ \left. \left. \gamma  \left( |\partial_y \phi_1|^2 + |\partial_x \phi_2|^2 \right) \right]  
+ U(\rho_1, \rho_2) +{\cal O}(\partial^4) 
\right\} .
\label{eq:EFTspiral} 
\end{eqnarray} 
To quartic order in the fields, 
the potential term $U(\rho_1, \rho_2)$ takes the form 
\begin{eqnarray} 
\textstyle U(\rho_1, \rho_2) = \frac{u_1}{2} \left(\rho_1 + \rho_2 -\rho_0 \right) ^2 + u_2 \rho_1 \rho_2 . 
\end{eqnarray} 
The latter can be viewed as simply a rewriting of
Eq.~\eqref{eq:spiralenergy}, 
and 
{the former following the standard notation~\cite{2002_Berges_ERG_PHYSReport}} allows
for slow spatial variations of the spiral order parameters.   
The theory thus has a momentum cutoff $\Lambda$. 
Symmetries  are  made transparent in the rewriting form of the free energy. 
Note that both axial and diagonal spiral phases are described by the
same free energy, up to a $\pi/4$ rotation of coordinates.

Introducing  dimensionless parameters 
$
\tilde{u}_{1,2} = Z^{-2} \Lambda^{-2} u_{1,2} \,, 
\tilde{\rho}_0 = Z \rho_0$, 
from the free energy in Eq.~\eqref{eq:EFTspiral}, 
  calculations of $1$PI RG equations are 
standard following Ref.~\cite{2002_Berges_ERG_PHYSReport}, 
and two illuminating limits are given here. First, 
as $\tilde{\rho}_0$ approaches  $\infty$, $\beta$-functions are asymptotically  
\be 
\textstyle \beta (\tilde{\rho}_0) =   {\cal O} (\tilde{\rho}_0 ^{-2}), \, \,\, 
\eta = -\Lambda \partial_\Lambda \log Z  =  \frac{1}{4 \pi \sqrt{\gamma} \tilde{\rho}_0} + {\cal O} (\tilde{\rho}_0 ^{-2}) , \nn
\ee 
\begin{eqnarray}
 \textstyle \beta (\tilde{u}_1) &=& \textstyle (2-2\eta) \tilde{u}_1 - \frac{\log 2 }{2 \pi \sqrt{\gamma}} \tilde{u}_1 ^2 + {\cal O} (\tilde{\rho}_0 ^{-3}),  \nn \\
\textstyle \beta (\tilde{u}_2 ) &=& \textstyle (2-2\eta) \tilde{u}_2 - \frac{\log 2}{2\pi \sqrt{\gamma}} \tilde{u}_1 \tilde{u}_2 + {\cal O} (\tilde{\rho}_0 ^{-3}).
\label{eq:rhoinflimit}
\end{eqnarray} 
We thus have an approximate fixed line at $\tilde{u}_1 ^* = \frac{4 \pi \sqrt{\gamma} (1-\eta^*) }{\log 2 }$, 
with $\eta ^* = \frac{1}{4\pi \sqrt{\gamma} \tilde{\rho}_0} $. This fixed line is parametrized by $\tilde{\rho}_0$. 
The second limit is $\tilde{\rho}_0 = 0, \gamma \to 1$, where $\beta$-functions are 
\begin{eqnarray} 
\textstyle \beta (\tilde{u}_1) &=& \textstyle 2 \tilde{u}_1 - \frac{\log 2 }{\pi \sqrt{\gamma}} \left\{ 5 \tilde{u}_1 ^2 + (\tilde{u}_1 + \tilde{u}_2) ^2 \right\}, \nn \\ 
\textstyle \beta (\tilde{u}_2) &=& \textstyle 2 \tilde{u}_2 - \frac{\log 2} {\pi \sqrt{\gamma}} \left[ 6 \tilde{u}_1 + \tilde{u}_2 \right]\tilde{u}_2  + {\cal O} ((\gamma-1)^2). 
\label{eq:rho0limit} 
\end{eqnarray} 
Here we have $\beta(\tilde{u}_1) = \beta(\tilde{u}_2) =0$ at the fixed points $(\tilde{u}_1 ^* = \frac{\pi \sqrt{\gamma}}{3 \log 2}, \tilde{u}_2 ^* =0)$. 

\begin{figure}[htp]
\includegraphics[bb = 1 3 585 221,angle=0,width=.9\linewidth]{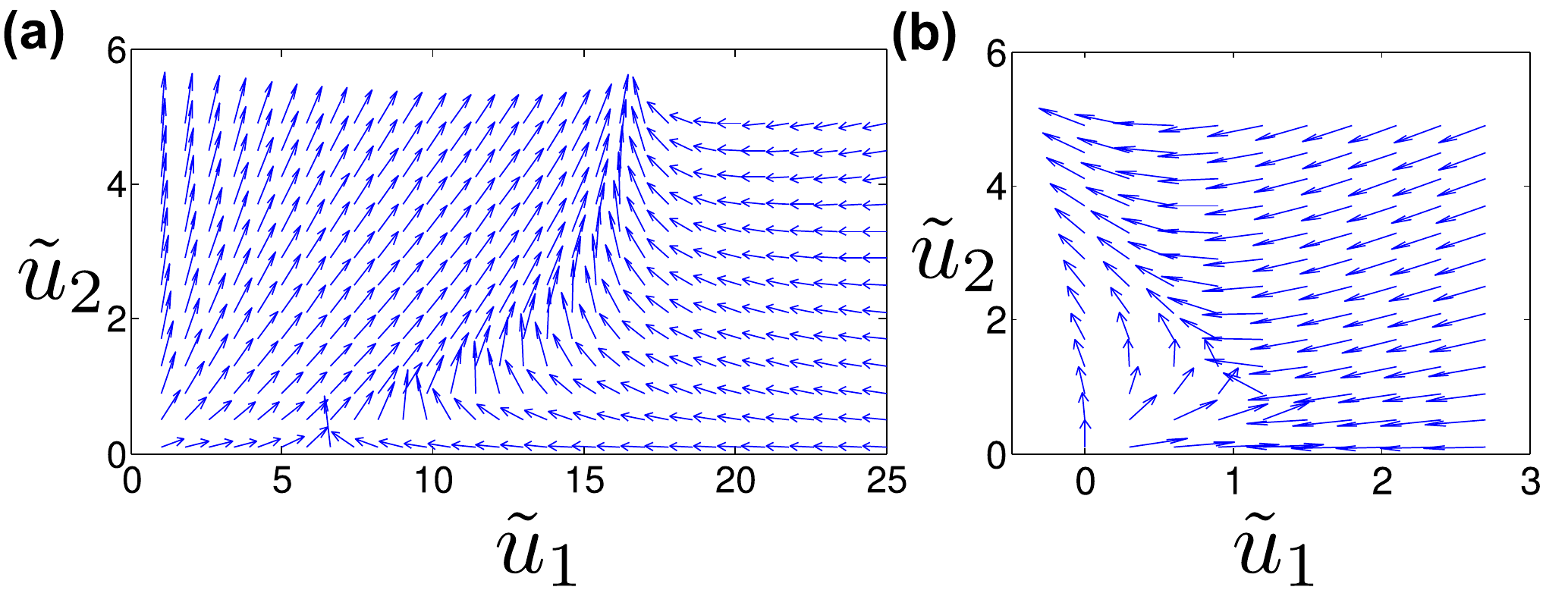}
\caption{ Vector illustration of  $(\beta (\tilde{u}_1), \beta (\tilde{u}_2))$.  In this plot we use $\gamma =1$. 
In (a) where $\tilde{\rho}_0 =1$, there is an approximate fixed point at 
$\left( \tilde{u}_1 ^* , \tilde{u}_2 \to +\infty\right)$. In (b), where $\tilde{\rho}_0 =0$, 
RG would flow to a regime with $\tilde{u}_1 <0$. 
For $\gamma$ slightly deviated from $1$, e.g., $\gamma = 1.5$, the qualitative features of this plot do not change. 
}
\label{fig:dg1dg2}
\end{figure}

In the intermediate regime 
we 
solve RG equations numerically and  
find that $\tilde{\rho}_0$ flows to zero above the  KT transition of spiral states 
when $\tilde{\rho}_0$ is smaller than some critical value $\tilde{\rho}_0 ^c$, 
dependent on $\tilde{u}_1$ and $\tilde{u}_2$. 
Thus we know from the limit (Eq.~\eqref{eq:rho0limit} and Fig.~\ref{fig:dg1dg2}~(b)),  $\tilde{u}_1$ flows to 
negative values and $\tilde{u}_2>0$ flows to large positive values above the KT transition, which predicts 
a nematic phase with an order parameter $\langle \rho_1 - \rho_2\rangle$. 
This indicates the nematic phase melts to paramagnetic  
through a first order phase transition at higher temperature.

}

\noindent{\bf Conclusion.}    
We determined a magnetic phase diagram for 2d electron systems with
weak but non-negligible SOC, relevant to oxide
heterostructures, but which is largely independent of the still
unresolved microscopic origin of magnetism.  Remarkably, we found a
skyrmion lattice phase similar to recent observations in helimagnets.
The complex spin textures and nematic state we found should be
detectable magnetically (see e.g. the discussion of magnetization of
spiral states in Ref.\cite{banerjee2013ferromagnetic}) but also
through transport, which should evince spontaneous anisotropy as well
as non-linear effects typical in sliding incommensurate
spin-density-waves.  The influence of Berry phases on electrons~\cite{2009_Yi_Nagaosa_PRB},
e.g. the anomalous Hall effect, is a promising direction for the
future.  These states may also be relevant to certain schemes for
engineering Majorana fermions, which require non-collinear magnetic
moments\cite{2013_Bernevig_Majorana_arXiv}.

\paragraph*{\bf Acknowledgement. }
We thank Eun-Gook Moon, Jeremy Levy, Kai Sun for helpful discussions.
X.L. would like to thank KITP, UCSB for hospitality and support by NSF
PHY11-25915.  The work at Pittsburgh is supported by A. W. Mellon
Fellowship (X.L.), AFOSR (FA9550-12-1-0079), ARO (W911NF-11-1-0230) and DARPA
OLE Program through ARO (X.L. and W.V.L.).  L.B. was supported by NSF
grant DMR-12-06809.

\bibliography{mag_oxide_interface}
\bibliographystyle{apsrev}

\begin{widetext} 

\newpage 

\renewcommand{\thesection}{S-\arabic{section}}
\renewcommand{\theequation}{S\arabic{equation}}
\setcounter{equation}{0}  
\renewcommand{\thefigure}{S\arabic{figure}}
\setcounter{figure}{0}  

{\bf Spirals and skyrmions in two dimensional oxide heterostructures 
-- Supplementary Materials} 

\begin{figure}[htp]
\includegraphics[angle=0,width=.5\linewidth]{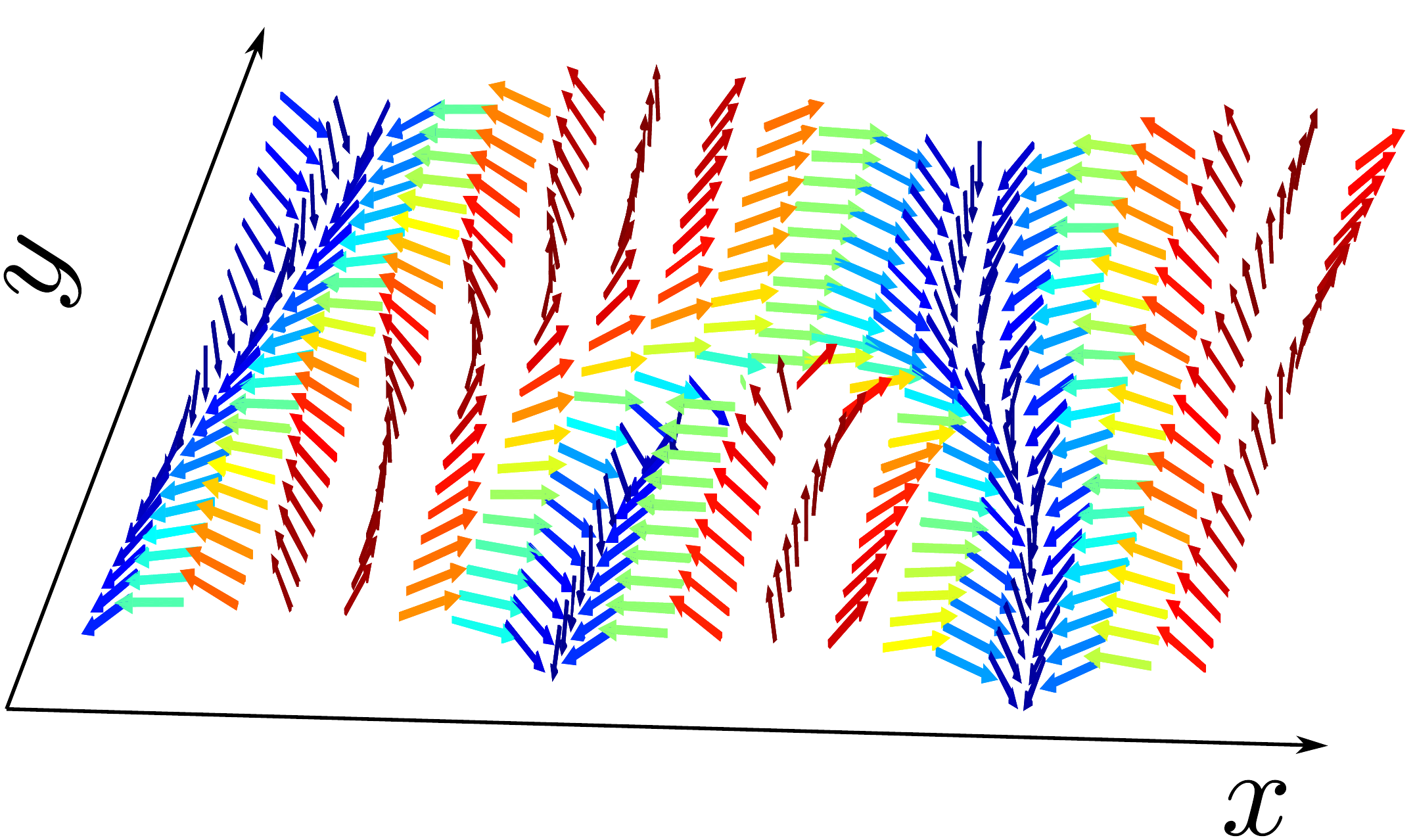}
\caption{ Spin configuration of a spiral state with one vortex. Spin moments in this spiral state without vortices 
vary in the $x$ direction and thus form stripes. The presence of a vortex here causes dislocation of the stripe order. Fluctuations of vortices/dislocations tend to restore 
the translation symmetry. 
}
\label{fig:singleqvortex}
\end{figure}

\begin{figure}[htp]
\includegraphics[angle=0,width=.5\linewidth]{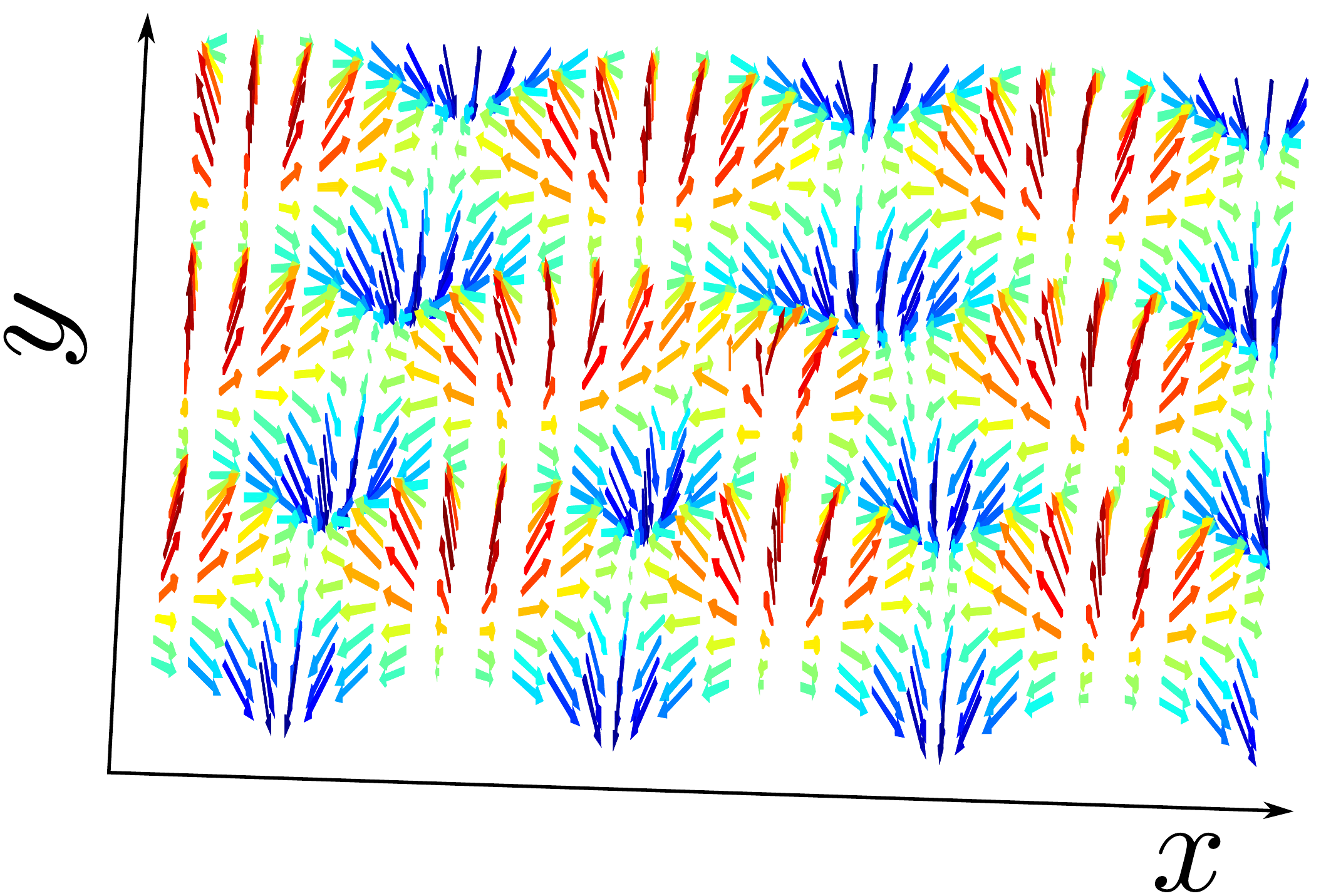}
\caption{ Spin configuration of a skyrmion lattice state with one vortex. The vortex here 
causes dislocation of the lattice, fluctuations of which tend to restore the translation symmetry. 
}
\label{fig:doubleqvortex}
\end{figure}

\end{widetext} 

\end{document}